\documentclass[aps,prl,tightenlines,floatfix,showpacs,twocolumn]{revtex4}
\usepackage{graphics}
\usepackage{bm}
\usepackage{amsmath}
\usepackage{amssymb}
\usepackage{txfonts} 
\usepackage{supertabular}
\usepackage{array}

\newcommand{\sfrac}[2]{\textstyle\frac{#1}{#2}}

\begin{document}

\title{Importance of resonance widths in low-energy scattering of
  weakly-bound light-mass nuclei}

\author{P. R. Fraser$^{1}$}
\email{paul.fraser@curtin.edu.au}
\author{K. Massen-Hane$^{1}$}
\author{K. Amos$^{2,3}$}
\author{I.~Bray$^{1}$}
\author{L. Canton$^{4}$}
\author{R.~Fossi{\'o}n$^{5}$}
\author{A.~S.~Kadyrov$^{1}$}
\author{S.~Karataglidis$^{3,2}$}
\author{J.~P.~Svenne$^{6}$}
\author{D.~van~der~Knijff$^{2}$}

\affiliation{
$^{1}$ Department of Physics, Astronomy and Medical Radiation
Sciences, Curtin University, GPO Box U1987, Perth 6845, Australia\\
$^{2}$ School of Physics, University of Melbourne, Victoria 3010, Australia\\
$^{3}$ Department of Physics, University of Johannesburg, P.O. Box 524,
  Auckland Park, 2006, South Africa\\
$^{4}$ Istituto Nazionale di Fisica Nucleare, Sezione di Padova, I-35131, 
Italy\\
$^{5}$ Instituto de Ciencias Nucleares, Universidad Nacional
Aut\'onoma de M\'exico, 04510, Ciudad de M\'exico, Mexico\\
$^{6}$ Department of Physics and Astronomy, University of Manitoba and
  Winnipeg Institute for Theoretical Physics, Winnipeg, MB, R3T 2N2,
  Canada}

\date{\today}

\begin{abstract}
What effect do particle-emitting resonances have on the scattering
cross section? What physical considerations are necessary when
modelling these resonances? These questions are important when
theoretically describing scattering experiments with radioactive ion
beams which investigate the frontiers of the table of nuclides, far
from stability. Herein, a novel method is developed that describes
resonant nuclear scattering from which centroids and widths in the
compound nucleus are obtained when one of the interacting bodies has
particle unstable resonances.  The method gives cross sections without
unphysical behavior that is found if simple Lorentzian forms are used
to describe resonant target states.  The resultant cross sections
differ significantly from those obtained when the states in the
coupled channel calculations are taken to have zero width, and
compound-system resonances are better matched to observed values.
\end{abstract}

\pacs{24.10.Eq; 24.30-v; 25.40.Dn; 25.40.Fq; 25.60-t}

\maketitle

The advent of radioactive ion beams has allowed exploration of nuclei
far from the valley of stability, and has led to an immense
experimental
effort~\cite{Be88,Or11,Hu11,Be11,Ba11,Na11,Du11,Ba14,Ba14a,Be03,Le10,Ok12,Su03,Ra11}.
Theoretical studies of these systems is vital for interpretation of
the resultant data.  Elastic scattering of two nuclei at low energies
often gives cross sections displaying resonances associated with
properties of the compound system; the analysis of which is
appropriately done with a coupled-channel theory in which the
low-energy spectra of the nuclei concerned are most relevant in
defining the coupling interactions. Usually, however, those states are
not considered to be resonances. Herein we present results found using
a theory in which those target state resonance properties are taken
into account. As detailed below, this requires a
mathematically-robust, energy-dependent shape to avoid unphysical
behaviors in calculated observables, such as vanishing bound states,
irregular behavior at the scattering threshold, and with the
requirement of causality being restored.

To this end, a multi-channel algebraic scattering (MCAS)
method~\cite{Am03} is used. MCAS solves coupled-channel
Lippmann-Schwinger equations in momentum space using the
Hilbert-Schmidt expansion of amplitudes.  In this method, two-body
nuclear scattering potentials are expanded into a series of
sturmians~\cite{We65,Ra91,Am03}, and then a corollary between
separable scattering potentials and separable $T-$matrices of the
Lippmann-Schwinger equation delivers solutions without explicitly
solving the integral equations. Scattering potentials used for this
investigation treat the nuclear target as having collective rotor
character~\cite{Ta65}.  However, in order to account the coupling of
the incident nucleon to Pauli-forbidden orbits in the target states,
one must also include an orthogonalising pseudo-potential (OPP)
\cite{Kr74,Ku78,Sa69,Ca05,Am13}, which has also been used in atomic
physics \cite{Mi99,Iv03}.

By solving the Lippmann-Schwinger equations in momentum space, one may
describe within the same method both the bound (to particle emission)
and scattering states of the compound nucleus. The bound states may be
found by assuming negative projectile energies in the sturmian
equations, and at the corresponding binding energies the sturmian
eigenvalues are real and have a value of 1.  For positive energies, to
systematically identify all resonance structures we use a spectral
representation of the $S$-matrix in terms of complex sturmian
eigenvalues~\cite{We65}. The trajectories of the eigenvalues in the
complex-energy plane, in particular in the vicinity of the
pole-position $P(1,0)$, can be employed to determine each resonance
centroid and width contained in the $S$-matrix, no matter how narrow
or large the resonance may be~\cite{Am03}.

Separable sturmian expansions of the chosen interaction potential are
made using a finite ($n$) set of sturmians (${\hat \chi}_{cn}(p)$);
functions that are generated from the same interactions for each
channel ($c$), where $c$ denotes a unique set of quantum numbers.
Obtaining the sturmian eigenstates, $\eta_{p}$, requires specification
of the Green's function~\cite{Am03}
\begin{align}
  \left[ \mathbf{G}_0 \right]_{nn'} = &\mu \left[
  \sum^{\text{open}}_{c = 1} \int^{\infty}_0 \hat{\chi}_{cn}(x)
  \frac{x^2}{ k^2_c - x^2 + i\varepsilon } \hat{\chi}_{cn'}(x) \, dx
  \right. \nonumber\\
  &- \left. \sum^{\text{closed}}_{c = 1}
  \int^{\infty}_{0} \hat{\chi}_{cn}(x) \frac{ x^2 }{ h^2 + x^2 }
  \hat{\chi}_{cn'}(x) \, dx \right].
\label{Greens1}
\end{align}
where the wave numbers are
\begin{equation}
k_c = \sqrt{\mu(E - \varepsilon_c)} \: \text{\ and\ } \:
h_c = \sqrt{\mu(\varepsilon_c - E)}\; ,
\label{kandh1}
\end{equation}
$\varepsilon_c$ is the target-state centroid and $E$ is the projectile
energy. Typically, the Green's functions are solved by methods of complex
analysis.

The spectrum of the compound system is found from the resolvent in the
$T$-matrix, namely $\left[ \mbox{\boldmath $\eta$}-{\bf G}_0
  \right]^{-1}$ where $\left[{\mbox {\boldmath $\eta$ }}\right]_{nn'}
= \eta_n\ \delta_{nn'}$ with $\eta_n$ being the sturmian eigenvalues.
The bound states of the compound system are defined by the zeros of
that matrix determinant when the energy is $E < 0$; all channels then
being closed.

\begin{figure}[htp]
\begin{center}
\scalebox{0.38}{\includegraphics*{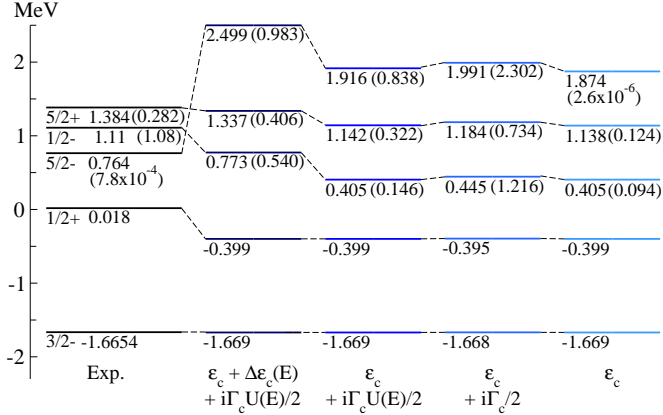}}
\end{center}
\caption{ \label{Be8+n-spec}(Color online.) Experimental spectrum of
  $^9$Be compared with MCAS calculations with target states defined as
  per labels (see text). Unbracketed numbers are excitation energies,
  bracketed numbers are widths, all in MeV.}
\end{figure}

\begin{figure}[htp]
\begin{center}
\scalebox{0.417}{\includegraphics*{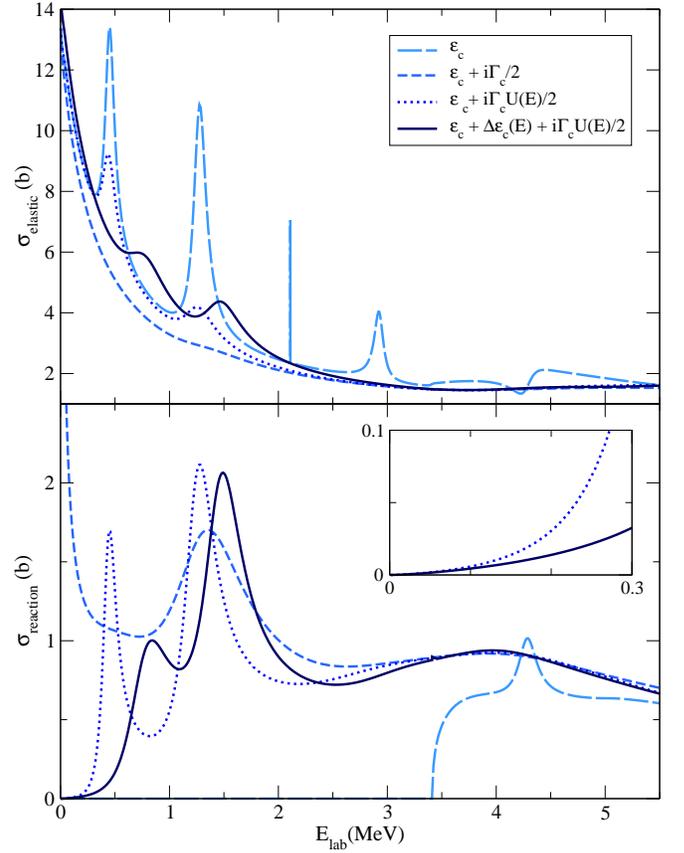}}
\end{center}
\caption{ \label{Be8+n-xsect}(Color online.) Calculated $n+^8$Be
  elastic scattering (top panel) and reaction (bottom panel) cross
  sections. Inset shows threshold behaviour of the reaction cross
  sections.}
\end{figure}

Results using the Green's function Eq.~(\ref{Greens1}) (and from those
later given) are shown in Figs.~\ref{Be8+n-spec} and
\ref{Be8+n-xsect}; the first presents spectra of ${}^9$Be as an
$n$+${}^8$Be cluster and the second a set of total elastic and
reaction cross sections in the energy range to just over 5
MeV. ${}^8$Be was treated as a rotor with quadrupole deformation and
three states of it, $0^+_{g.s.}$, $2^+_1$ and $4^+_1$, used in the
coupling.  In Fig.\ref{Be8+n-spec}, the spectrum for ${}^9$Be found
using Green's functions as per Eq.~(\ref{Greens1}) is shown in the
column furthest right.  For comparison, the experimental
spectrum~\cite{Ti04}, is shown on the far left. Fig.~\ref{Be8+n-xsect}
displays the cross sections found from the same calculation (and
others discussed later) whose spectrum is shown in
Fig.~\ref{Be8+n-spec}. The results are identified by the same
notation. In this case where Eq.~(\ref{Greens1}) is used, the reaction
cross section only becomes non-zero above the energy of the first
target state, which is at 4.81~MeV (lab frame), as necessary.

However, in the low-energy and low-mass regime where compound-system
resonances are important, it is appropriate to take
particle-instability of target states into account by modifying the
Green's functions. In its most basic form~\cite{Fr08a}, this is done by
adding a complex component to the target-state energy. That is, the
description of the target state energy becomes:
\begin{equation}
\varepsilon_c + i\sfrac{\Gamma_c}{2} \,,
\label{allwidths}
\end{equation}
and the Green's functions thus become
\begin{align}
  \left[ \mathbf{G}_0 \right]_{nn'} = &\mu \left[
  \sum^{\text{open}}_{c = 1} \int^{\infty}_0 \hat{\chi}_{cn}(x)
  \frac{ x^2 \left[ k_c^2 - x^2 - \sfrac{ i \mu \Gamma_c }{2} \right] }
  {\left[ k_c^2 - x^2 \right]^2 + \sfrac{ {\mu}^2 {\Gamma_c}^2 }{4}} 
  \hat{\chi}_{cn'}(x) \, dx
  \right. \nonumber\\
  &- \left. \sum^{\text{closed}}_{c = 1} 
  \int^{\infty}_{0} \hat{\chi}_{cn}(x) 
  \frac{ x^2 \left[ h_c^2 + x^2 + \sfrac{ i \mu \Gamma_c }{2} \right] }
  { \left[ h_c^2 + x^2 \right]^2 + \sfrac{ {\mu}^2 {\Gamma_c}^2 }{4} } 
  \hat{\chi}_{cn'}(x) \, dx \right].
\label{Greens2}
\end{align}
This equation has no poles on the real axis, and integration may
proceed normally~\cite{Fr08a}. The spectrum in Fig.~\ref{Be8+n-spec}
identified by the complex energy $\epsilon_c + i \frac{\Gamma}{2}$
resulted on using the same interaction as before but with the $2^+$
and $4^+$ states of $^8$Be having their known particle-emission
widths~\cite{Ti04}.

Cross sections are calculated using the $S$-matrix which has the general form:
\begin{multline}
  S_{cc'} = \delta_{cc'} \;-\\ i^{l_{c'} - l_c + 1} \pi \mu \sum^{N}_{n,n'
  = 1} \hspace{-2mm}\sqrt{k_c} \hat{\chi}_{cn}(k_c) \left( \left[ {\mbox{\boldmath
  $\eta$}} - \mathbf{G}_0 \right]^{-1} \right)_{nn'} \hat{\chi}_{c'n'}(k_{c'})
  \sqrt{k_{c'}} \,.
\label{Smatrix}
\end{multline}
where $\eta$ is an array of sturmian eigenvalues and $G_0$ are the
same Green's function used to defined in Eq.~(\ref{Greens1}) when no
target state widths are considered, and Eq.~(\ref{Greens2}) with
states described as per Eq.~(\ref{allwidths}).
As the systems considered herein do not have
particle emission widths in their ground states, the sturmians ``in
the elastic channel" $\hat{\chi}_{1n}(k_1)$ and $\hat{\chi}_{1n}(k_1)$
will not be different from cases where no target-state widths are
considered.  However, $S$-matrices and thus cross sections will still
be altered by the inclusion of particle-emission widths through the
channels of $\left( \left[ {\mbox{\boldmath $\eta$}} - \mathbf{G}_0
  \right]^{-1} \right)_{nn'}$ not involving the target ground state.
The cross sections that results from using complex energies for the
$2^+_1$ and $4^+_1$ states in ${}^8$Be, are shown in Fig. 2,
identified by the notation $\epsilon_c + i \frac{\Gamma}{2}$.

Of note, with particle-emission considered, the reaction cross section
is non-zero from zero projectile energy upwards, due to loss of flux
from target decay. However, their asymptotic behaviour at low
projectile energies is unphysical, and is due to the Lorentzian form
that implicitly defines the target states in Eq.~(\ref{Greens2}) being
non-zero at and below the scattering threshold (as also observed in a
technical note, Ref.~\cite{Ca11}). This also affects the energy of
bound states, causing them to become spuriously unstable.

To overcome this non-physical behaviour, a scaling factor is
applied to target-state widths, such that the target states are now
described as
\begin{equation}
\varepsilon_c + i\sfrac{U(E)\cdot\Gamma_c}{2} \,,
\label{Uwidths}
\end{equation}
which changes the Green's functions of Eq.~(\ref{Greens2}) by
multiplication of $\Gamma$ in both integrals by $U(E)$. As minimum
conditions, the scaling function $U(E) = 0$ when $E \le 0$, $U(E) = 1$
at energy centroid, and $U(E) \rightarrow 0$ as $E \rightarrow
\infty$. In addition, to fully eliminate asymptotic behaviour in the
reaction cross sections as $E \rightarrow +0$, it is required that
$\frac{dU(E)}{dE} \rightarrow 0$ as $E \rightarrow +0$.  See
Ref.~\cite{Ca11} for an example of where a scaling function was
investigated where the last condition was not met (and where causality
correction, discussed below, was not addressed.)

The concept of energy dependent widths goes back to
Wigner~\cite{Te52}, and is widely used in nuclear cross section
estimates~\cite{Br12}.  Typically, the low-energy dependence of such
scaled resonances are ruled by the centrifugal (and eventually
Coulomb) barrier. The probability of formation of a resonance is
modulated at low energies by these ``penetration" factors.  It is
these factors which lead to the requirement on the scaling functions
that they and their derivatives are vanishing at the scattering
threshold.

However, the introduction of energy-dependent widths necessitates an
energy-dependent addition to the target-state centroid, transforming
the energy of the state \textit{viz.}
\begin{equation}
\varepsilon_c + \Delta \varepsilon_c(E) + i\sfrac{\Gamma_c \cdot U(E)}{2} \,.
\label{fullstate}
\end{equation}
This is because the Green's functions define the sturmian eigenvalues
of the expansion of the potential. Thus, making the prescription of
the target states complex in effect makes the potential an optical
potential. As detailed in Refs.~\cite{Co62,Ma86}, energy-dependent
complex-components in optical potentials lead to a wave equation that
violates causality unless the potential is restricted by the addition
of a dispersion relation to the real part of the potential. These
concepts have been used in phenomenological optical models in, for
example, Refs.~\cite{Ma86a,Ca91}.

Here, the dispersion relation is an energy-dependent adjustment
of the target-state centroid energy, $\Delta \varepsilon_c(E)$, given by
\begin{equation}
\Delta \varepsilon_c(E) = \frac{\Gamma_c}{2} \, \frac{1}{\pi} \, 
                P \int^\infty_0\frac{U(E')}{E'-E}dE' \ .
\label{Deltae}
\end{equation}
This manifests in Eq.~(\ref{Greens2}) (with $\Gamma$ multiplied by
$U(E)$) as wave numbers with the form
\begin{equation}
k_c = \sqrt{\mu(E - \varepsilon_c - \Delta \varepsilon_c(E))} 
\: \text{\ and\ } \:
h_c = \sqrt{\mu(\varepsilon_c + \Delta \varepsilon_c(E) - E)}\; .
\label{kandh2}
\end{equation}
Many nuclear targets have a ground state with no particle-emission
width, and when considering the channels involving those ground
states, the wave numbers have the form of Eq.~(\ref{kandh1}) rather
than Eq.~(\ref{kandh2}), and the Green's function of
Eq.~(\ref{Greens1}) applies rather than Eq.~(\ref{Greens2}) (modified
by $U(E)$ and $\Delta \varepsilon_c(E)$).

\begin{figure}[htp]
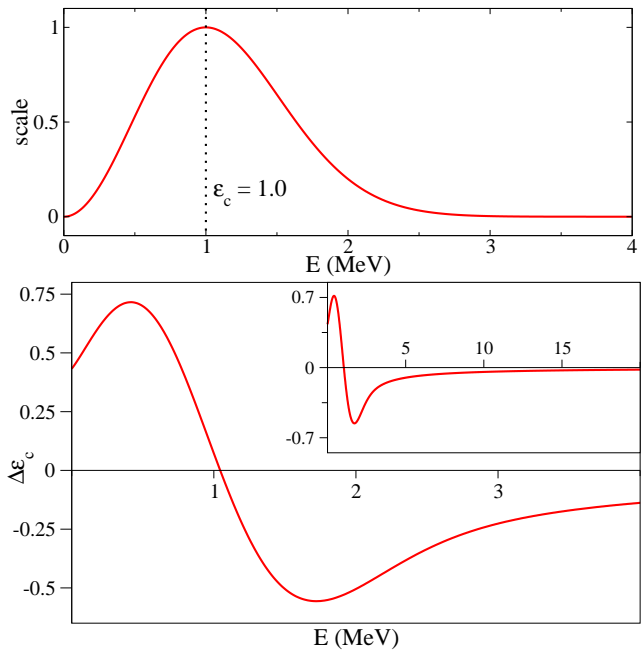

\begin{center}
\scalebox{0.35}{\includegraphics*{wigner.eps}}
\scalebox{0.35}{\includegraphics*{dtargex.eps}}
\end{center}
\caption{ \label{ddtargex}(Color online.) Top: Scaling function $U(E)$ of
  Eq.~(\ref{Wigner}) for $q=1$, and $Z=2$ with
  $\varepsilon_c=1$. Bottom: Numerically evaluated $\Delta
  \varepsilon_c(E)$ with $U(E)$ having parameters as above and
  $\Gamma_c=2$ MeV. Insert shows approach to -0 as projectile energy
  increases.}
\end{figure}

One candidate for an energy-dependent target-state width scaling is
based upon a Wigner distribution~\cite{Me91}, modified to meet the
necessary conditions:
\begin{equation}
U(E) 
= e^{q}\left(\frac{E}{\varepsilon_c}\right)^Ze^{-q\left({E}/{\varepsilon_c}\right)^Z}
       H(E)
\label{Wigner}
\end{equation}
where $q$ and $Z$ are positive parameters. The Heaviside function
ensures proper bound-state properties. The upper panel
of Fig.~\ref{ddtargex} shows Eq.~(\ref{Wigner}) for $q=1$, and
$Z=2$. The lower panel shows the integrated result of
Eq.~(\ref{Deltae}) with $\varepsilon_c =1$ MeV and $\Gamma_c = 2$ MeV.

At projectile energies below that of a resonant target state's actual
centroid, the effect of reducing the width of that state increases the
centroid used for purposes of defining the wave number. Conversely, at
projectile energies above the actual centroid, the reduction in target
state width decreases the centroid used. The transition from positive
to negative centroid correction occurs at $E < \varepsilon_c$ for
these values of $q$ and $Z$, which is caused by the exponential
suppression of the scaling function $U(E)$ at energies larger than
$E$.  As projectile energy tends to infinity, the centroid correction
tends to +0.

Column 2 of Fig.~\ref{Be8+n-spec} shows the resonances and bound
states of an MCAS calculation with resonant states defined as per
Eq.~(\ref{fullstate}), using the Green's function 
defined accordingly, and is labelled appropriately.
The calculation used the same potential as all the others, and in fact
the parameters were tuned for this case. The appropriately-labelled
curves in Fig.~\ref{Be8+n-xsect} show the resultant elastic and
reaction cross sections.  Column 3 of Fig.~\ref{Be8+n-spec} and the
matching curves of Fig.~\ref{Be8+n-xsect} show the results of the
energy-dependent scaling of widths but neglecting the causality
correction to the centroid energy.

It is seen from differences between columns 2 and 5 of
Fig.~\ref{Be8+n-spec} that consideration of nuclear instability in
scattering calculations has non-trivial impact upon compound-state
centroids, affecting how scattering potentials must be defined to
match experiment. The differences between column
2 and 3 show that the causality correction accounts for a significant
part of this variation. The result of the full physical description of
target states [column 2] gives the best centroid values for the
$\sfrac{1}{2}^-$ and $\sfrac{5}{2}^+$ resonances, the features that
dominate the calculated cross sections.

The $\sfrac{1}{2}^-$ resonance is only known to decay by neutron
emission, and the $\sfrac{5}{2}^+$ resonance by neutron and $\gamma$
emission~\cite{Ti04}, and so this MCAS calculation considers all
important components of the resonances' widths. The calculation with
no consideration of $^8$Be decay widths [column 5] leads a width for
the $\sfrac{1}{2}^-$ state that is only 9\% of that observed
experimentally, where the calculation with target-state width scaling
and causality correction [column 2] gives a result that is 72\% of
the known value. The calculation in which decay widths are included
but not scaled [column 4] produces 112\% of the known result, but as
with column 3, the calculation is unphysical as previously
discussed. Regarding the $\sfrac{5}{2}^+$ resonance, the width result
in column 5 is 44\% of the experimental value, while that of column 2
is 144\%, a slightly better ratio, and that of column 4 is a large
overestimation at 260\%. (The centroid of the $\sfrac{5}{2}^-$
resonance is poorly recreated in all calculations, and concordantly
the widths are over- or under-estimated in all cases by orders of
magnitude. This state is known to decay by $\alpha$ emission, not
considered here, as well as $n$ and $\gamma$ emission.) Thus,
certainly in the case of the $\sfrac{1}{2}^-$ resonance, and arguably
that of the $\sfrac{5}{2}^+$ resonance, consideration of particle
emission from target states is seen to be a necessary ingredient in
better describing scattering involving loosely-bound nuclei. Further
investigation of scaling factor forms may yield yet better
descriptions of compound-system resonance shapes.

The appropriately-labelled curves in Fig.~\ref{Be8+n-xsect} show cross
sections resulting from defining target states as per
Eq.~(\ref{fullstate}), and with target state width scaling but
neglecting the causality correction. Again the reaction cross section
is non-zero from zero projectile energy upwards due to flux loss, but
it is observed that the scaling factor successfully eliminates the
erroneous asymptotic rise in the reaction cross section near the
threshold. This is highlighted by the inset panel. Causality
restoration, by altering centroids, affects the shape of the cross
sections, with consequences for scattering-potential parameterisation.

\begin{figure}[t]
\begin{center}
\scalebox{0.45}{\includegraphics*{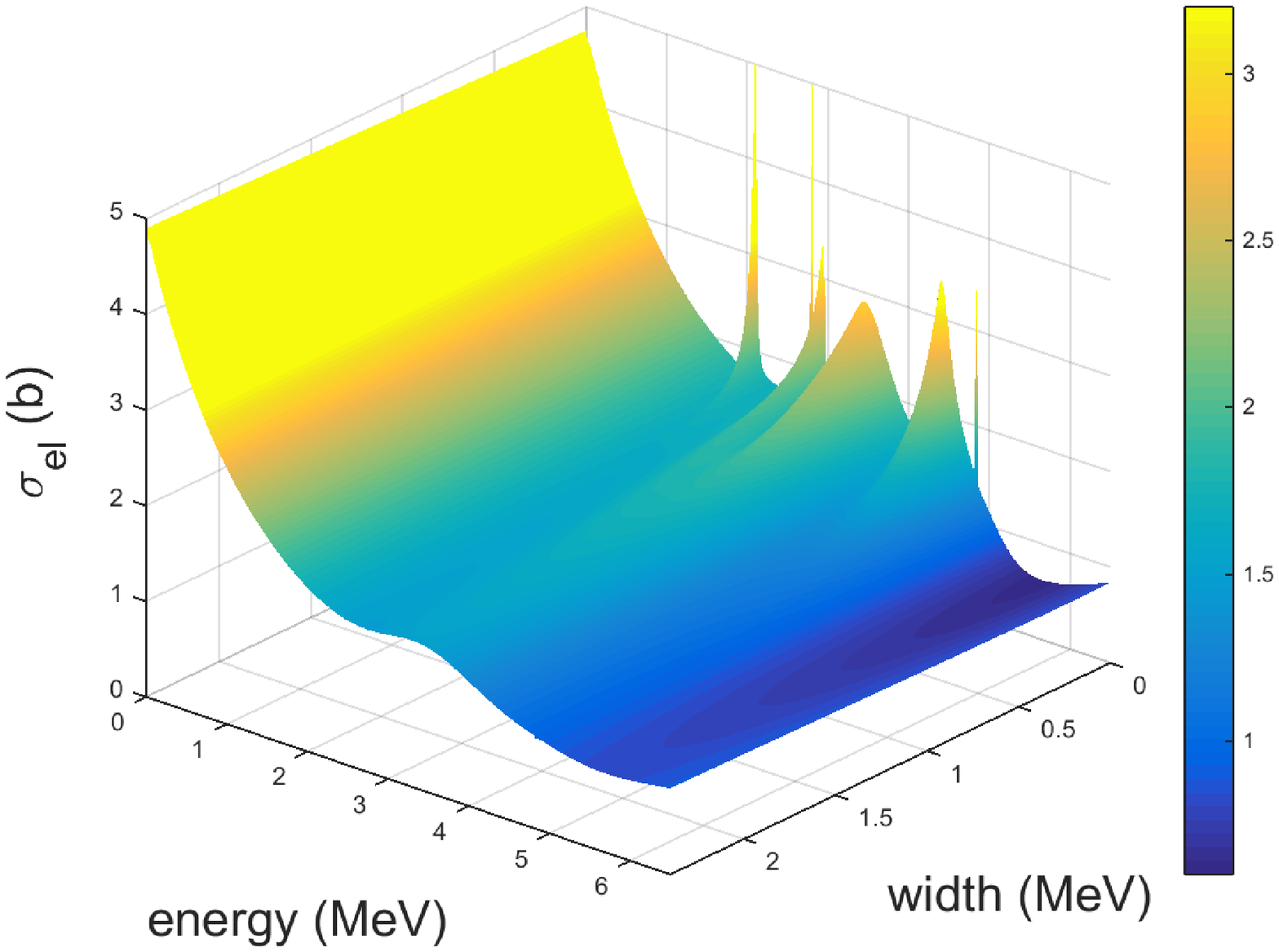}}
\scalebox{0.45}{\includegraphics*{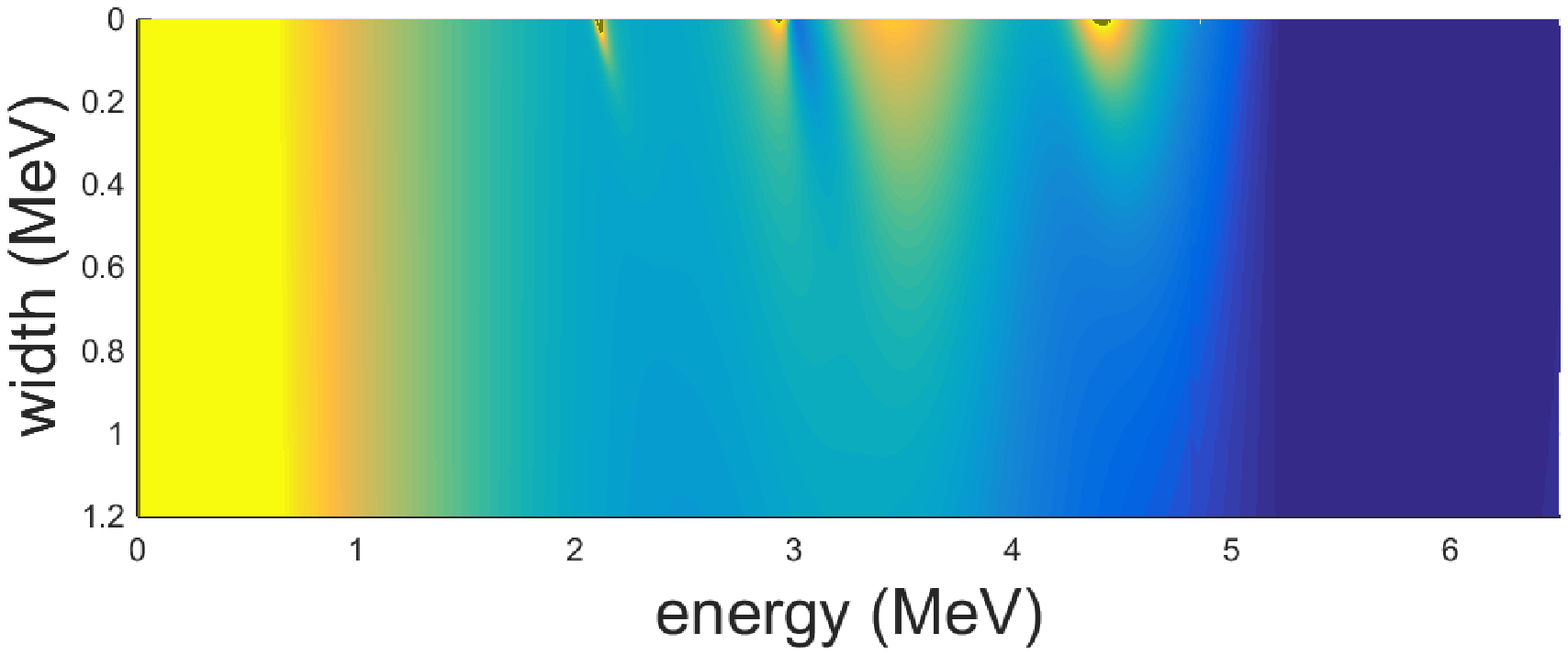}}
\end{center}
\caption{ \label{Wigner-elastic}(Color online.) Top: $n+^{12}$C
  elastic scattering cross section with \textit{gedanken}
  particle-emission widths, $\Gamma_c$, of $^{12}$C $2^+_1$ and
  $0^+_2$ states as per the axis. $E$ is the projectile
  energy. Bottom: contour map detail of the top panel. Target states
  are as per the right of Eq.~(\ref{fullstate}), with $q=1$, $Z=2$ in
  Eq.~(\ref{Wigner}).}
\end{figure}

\begin{figure}[htp]
\begin{center}
\scalebox{0.45}{\includegraphics*{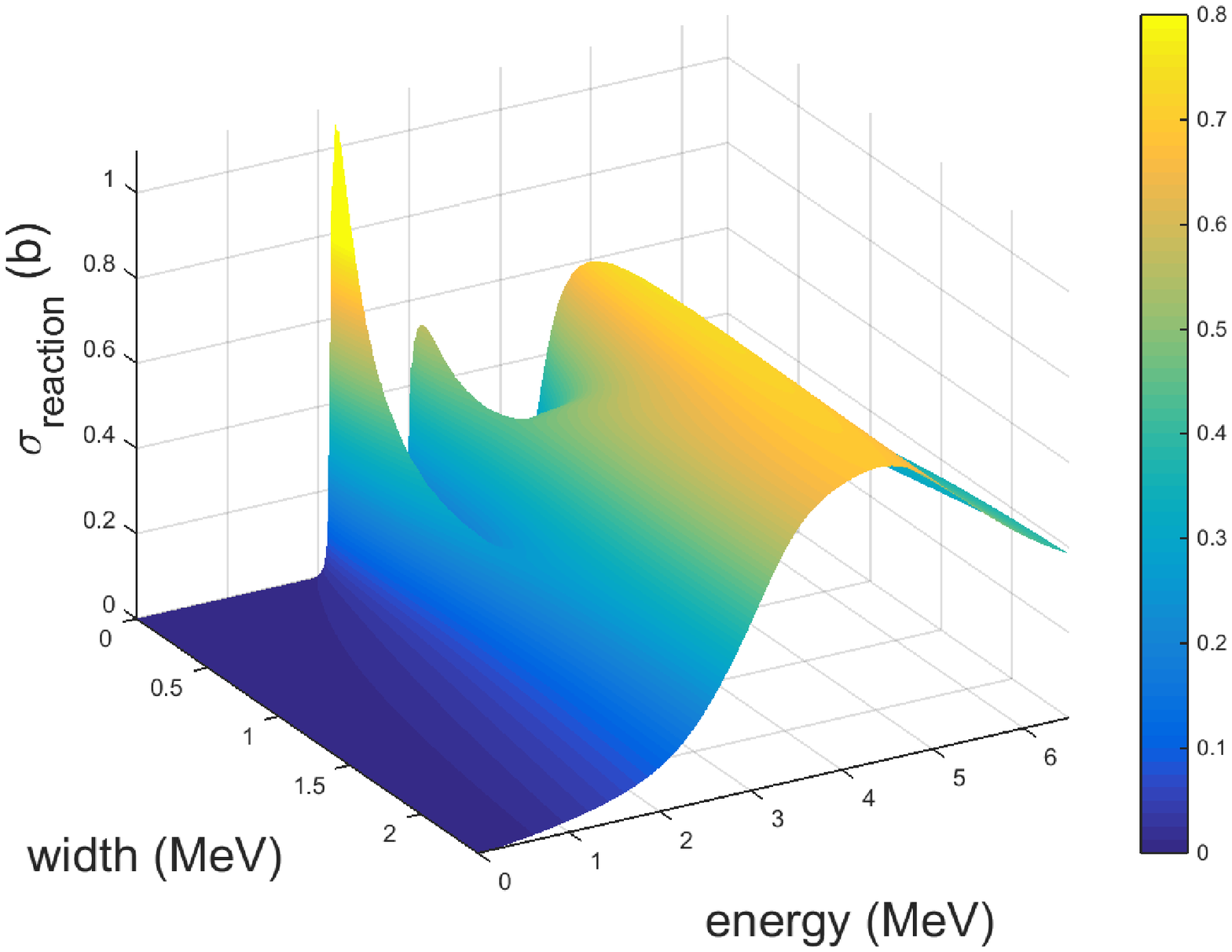}}
\scalebox{0.45}{\includegraphics*{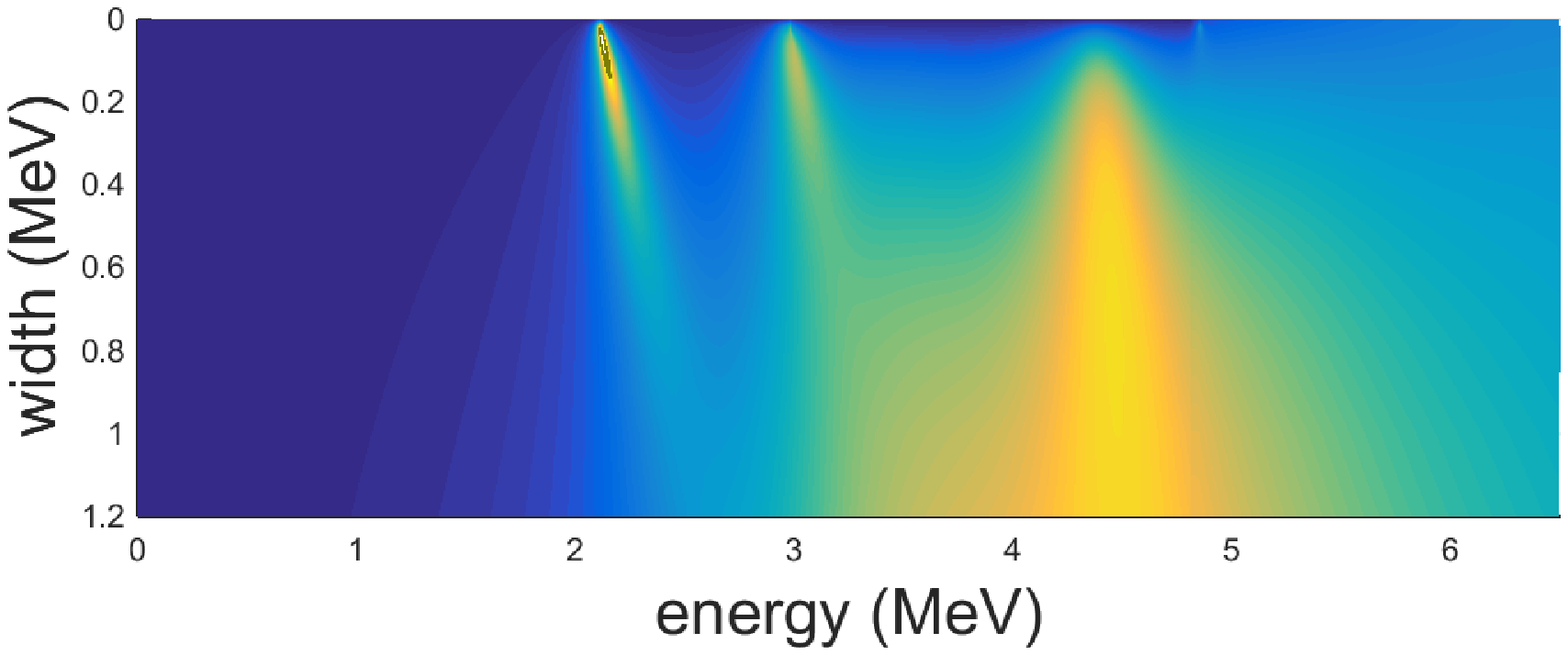}}
\end{center}
\caption{ \label{Wigner-reaction}(Color online.) Top: $n+^{12}$C
  reaction cross section with \textit{gedanken} particle-emission
  widths, $\Gamma_c$, of $^{12}$C $2^+_1$ and $0^+_2$ states as per
  the axis. $E$ is the projectile energy. Bottom: contour map detail
  of the top panel. Target states are as per the right of
  Eq.~(\ref{fullstate}), with $q=1$, $Z=2$ in Eq.~(\ref{Wigner}).}
\end{figure}

To further illustrate the effect of the Green's functions of
Eq.~(\ref{Greens2}) modified with $U(E)$ and $\Delta \varepsilon_c(E)$,
we examine a \textit{gedanken} case of the scattering of low-energy
neutrons from $^{12}$C, with coupling of the neutron to the $0^+_1$,
$2^+_1$ and $0^+_2$ states of $^{12}$C.  A range of artificial
particle-emission widths are assigned to the $2^+_1$ and $0^+_2$
states of the target, with the resulting elastic-scattering and
reaction cross sections shown in Figs.~\ref{Wigner-elastic} and
\ref{Wigner-reaction}, respectively.

Fig.~\ref{Wigner-elastic} shows that, while the inclusion of target
state widths has minimal impact upon the scattering background, with
increasing target-state widths, compound-system resonances are reduced
in amplitude and increased in width.  With increasing target-state
widths, narrow resonances are subsumed into the scattering
background. The wider compound-state resonances persist to greater
widths.  Note that, even when not discernible from the scattering
background, the method of obtaining resonances outlined above still
identifies them.


The second panel of Fig.~\ref{Wigner-reaction} shows that when
target-state widths equal 0~MeV, the reaction cross section only
becomes non-zero above the energy of the first target state, at 4.81
MeV (lab frame), as is necessary.  When target-state widths are
increased, the reaction cross section becomes non-zero for all
projectile energies greater than the scattering threshold, as particle
decay leads to flux loss. As target widths increase from zero,
compound-system resonances immediately appear and dominate this region
below the first target state energy, and then rapidly become subsumed
into the scattering background. No unphysical asymptotic behavior is
observed at projectile energies near the scattering threshold.

\begin{figure}[htp]
\begin{center}
\scalebox{0.417}{\includegraphics*{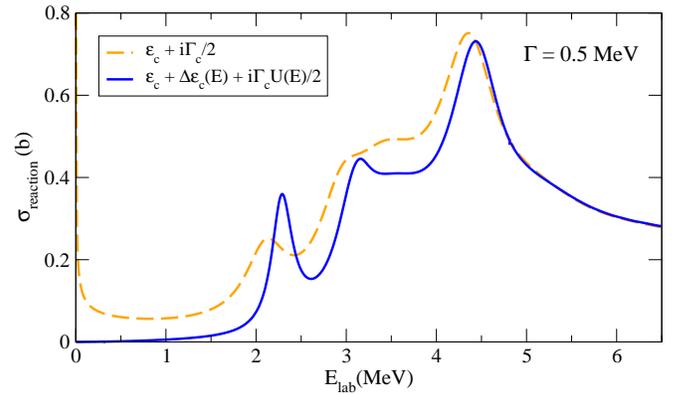}}
\end{center}
\caption{ \label{C12+n-xsect}Calculated $n+^{12}$C
  reaction cross for $\Gamma_c$~=~0.5 MeV.}
\end{figure}

To further examine behaviour of the reaction cross section near the
scattering threshold, Fig.~\ref{C12+n-xsect} shows the case of the
$^{12}$C $2^+_1$ and $0^+_2$ states each having a width of 0.5 MeV
with target states defined as per Eq.~(\ref{allwidths}) and as defined
by Eq.~(\ref{fullstate}) [being a cross section of
  Fig.~\ref{Wigner-reaction}]. As in the $^9$Be investigation, the
former has erroneous asymptotic behaviour as $E \rightarrow +0$, which
is eliminated in the latter.

In conclusion, a method of accounting for states that are
particle-unstable in nuclei undergoing low-energy resonant scattering
is developed, which is free of unphysical behaviour at the scattering
threshold and conserves causality. This is performed by choosing an
appropriate target-state resonance shape, modifying a Lorentzian by
use of widths dependent on projectile energy, with a correction to
target-state centroid energy. Resultant scattering cross sections are
markedly different from those found when particle instability is not
considered. Compound-system resonances decrease in magnitude and
increase in width, with otherwise narrow resonances becoming obscured
into the scattering background. This was shown to improve agreement
between calculated and observed widths of such resonances.  When using
parameter-driven scattering potentials, the effects of the
target-state resonance shape - and in the case energy-dependent
modified Lorentzians, the centroid correction - are non-trivial in
defining the potential.  Compound spectra associated with, and
scattering cross sections from, weakly-bound radioactive ion beams
with light-mass targets should be influenced by such considerations as
these.

This work is supported by the Australian Research Council, National
Research Foundation of South Africa and U.S. National Science
Foundation under Award No. PHY-1415656.

\bibliography{Fr15b}

\end{document}